\documentclass{ws-p8-50x6-00}
\begin{document}
%                   G. 't Hooft  macros version 2000
\newcounter{L1}

\newread\testifexists
\def\GetIfExists #1 {\immediate\openin\testifexists=#1
    \ifeof\testifexists\immediate\closein\testifexists\else
    \immediate\closein\testifexists\input #1\fi}
\def\epsffile#1{}

\immediate\openin\testifexists=e
    \ifeof\testifexists\immediate\closein\testifexists\else
    \immediate\closein\testifexists\input e\fipsf

%\magnification= \magstep1  % or use \magstephalf
%\tolerance=1600
%\parskip=5pt
%\baselineskip= 5 true mm \mathsurround=1pt
% \hsize=5.4in
% \vsize=6.6in % for screen
\font\smallrm=cmr8 \font\smallit=cmti8 \font\smallbf=cmbx8
\font\medrm=cmr9 \font\medit=cmti9 \font\medbf=cmbx9
\font\bigbf=cmbx12 \font\bigit=cmti12
    \def\Bbb#1{\setbox0=\hbox{$\tt #1$}  \copy0\kern-\wd0\kern .1em\copy0}
    \immediate\openin\testifexists=a
    \ifeof\testifexists\immediate\closein\testifexists\else
    \immediate\closein\testifexists\input a\fimssym.def %% for \Bbb A - Z %%

\def\secbreak{\vskip12pt plus .6in \penalty-200\vskip -2pt plus -.4in}
\def\prefbreak#1{\vskip#1\penalty-200\vskip-#1\vskip -5 true mm } %\prefbreak{distance}
\def\ref#1{${\,}^{\hbox{\smallrm #1}}$}
   \def\newsect#1{\secbreak\noindent{\bf #1}\medskip}
   \def\br{\vfil\break}\def\rechts#1{\hbox{\ }\hfil#1\break}
\def\hugeskip{\vskip12mm plus 3mm}
\def\Narrower{\par\narrower\noindent}   % never again use TeX's awful \narrower
\def\Endnarrower{\par\leftskip=0pt \rightskip=0pt}
\def\br{\hfil\break}    \def\ra{\rightarrow}

\def\cl{\centerline}
\def\low#1{_{\vphantom{Q}#1}}\def\high#1{^{\vphantom{\big[}#1}}
\def\ni{\noindent}      \def\pa{\partial}   \def\dd{{\rm d}}
\def\tl{\tilde}         \def\bra{\langle}   \def\ket{\rangle}

\def\a{\alpha}      \def\b{\beta}   \def\g{\gamma}      \def\G{\Gamma}
\def\d{\delta}      \def\D{\Delta}  \def\e{\varepsilon}
\def\h{\eta}         \def\k{\kappa} \def\l{\lambda}     \def\L{\Lambda}
\def\m{\mu}         \def\f{\phi}    \def\F{\Phi}        \def\vv{\varphi}
\def\n{\nu}         \def\j{\psi}    \def\J{\Psi}
\def\r{\varrho}     \def\s{\sigma}  \def\SS{\Sigma}
\def\t{\tau}        \def\th{\theta} \def\thh{\vartheta} \def\Th{\Theta}
\def\x{\xi}         \def\X{\Xi}     \def\ch{\chi}
\def\w{\omega}      \def\W{\Omega}  \def\z{\zeta}

\def\HH{{\cal H}} \def\LL{{\cal L}} \def\OO{{\cal O}}

\def\fn#1{\ifcase\noteno\def\fnchr{*}\or\def\fnchr{\dagger}\or\def
    \fnchr{\ddagger}\or\def\fnchr{\medrm\S}\or\def\fnchr{\|}\or\def
    \fnchr{\medrm\P}\fi\footnote{$^{\fnchr}$}
    {\scrunch#1\toe}\ifnum\noteno>4\global\advance\noteno by-6\fi
    \global\advance\noteno by 1}
    \def\scrunch{\baselineskip=11 pt \medrm}
    \def\toe{\vphantom{$p_\big($}}
    \newcount\noteno
    %    footnote with alternating symbol   %
\def\qu{\ {\buildrel {\displaystyle ?} \over =}\ }
\def\deff{\ {\buildrel{\rm def}\over{=}}\ }

\def\ddt{{{\rm d}\over {\rm d}t}}
\def\ffract#1#2{{\textstyle{#1\over#2}}}

\def\fract#1#2{{\textstyle{#1\over#2}}}

\def\half{\ffract12} \def\quart{\ffract14}

\def\part#1#2{{\partial#1\over\partial#2}}
\def\on{\over\displaystyle} \def\ref#1{${\vphantom{)}}^#1$}
\def\ex#1{e^{\textstyle#1}}

\def\bbf#1{\setbox0=\hbox{$#1$} \kern-.025em\copy0\kern-\wd0
        \kern.05em\copy0\kern-\wd0 \kern-.025em\raise.0433em\box0}
                % boldface in math mode.
\def\qu{\ {\buildrel {\displaystyle ?} \over =}\ }
\def\df{\ {\buildrel{\rm def}\over{=}}\ }

\def\dddot#1{{\buildrel{\displaystyle{\,...}}\over #1}}
\def\low#1{{\vphantom{]}}_{#1}}
 \def\bal{$\bullet$} \def\bel{$\circ$}
\def\ref#1{${\,}^{\hbox{\smallrm #1}}$}

\def\Gbar{\raise.13em\hbox{--}\kern-.35em G}
\def\lap{\setbox0=\hbox{$<$}\,\raise .25em\copy0\kern-\wd0\lower.25em\hbox{$\sim$}\,}
\def\glt{\setbox0=\hbox{$>$}\,\raise .25em\copy0\kern-\wd0\lower.25em\hbox{$<$}\,}
\def\gap{\setbox0=\hbox{$>$}\,\raise .25em\copy0\kern-\wd0\lower.25em\hbox{$\sim$}\,}
\def\nc{\medskip\noindent}
\def\iz{\quad = \quad}
\def\dys{\displaystyle}\def\scs{\scryptstyle}\def\txs{\textstyle}

%{\nopagenumbers %
%{\ }\vglue 1truecm \rightline{SPIN-2002/08}\rightline{ITF-2002/14}
%\rightline{hep-th/0204069} \hugeskip \cl{\bigbf

\title{{\rm\rightline{SPIN-2002/08}\rightline{ITF-2002/14}
\rightline{hep-th/0204069}}\br Large $N$}

%Tempe, Arizona, January 2002}
%\hugeskip

\author{Gerard 't Hooft }

\bigskip
\address{Institute for Theoretical Physics, Utrecht University,
Leuvenlaan 4 \\ 3584 CC Utrecht, the Netherlands \\ {\rm and} \\
%\smallskip
%\cl{and}
%\smallskip
Spinoza Institute, Postbox 80.195, 3508 TD Utrecht, the
Netherlands \\ E-mail: {\tt g.thooft@phys.uu.nl} \\ Internet: \tt
http://www.phys.uu.nl/\~{}thooft/}
%
%\hugeskip

\maketitle

\abstracts{ { \it
  Keynote Address,
presented at  the workshop on "The Phenomenology of Large $N_c$
QCD", Arizona State University, January, 2002. \\ \\ } In the
first part of this lecture, the $1/N$ expansion technique is
illustrated for the case of the large-$N$ sigma model. In
large-$N$ gauge theories, the $1/N$ expansion is tantamount to
sorting the Feynman diagrams according to their degree of
planarity, that is, the minimal genus of the plane onto which the
diagram can be mapped without any crossings.  This holds both for
the usual perturbative expansion with respect to powers of
${\tilde g}^2=g^2 N$, as well as for the expansion of lattice
theories in positive powers of $1/{\tilde g}^2$. If there were no
renormalization effects, the $\tilde g$ expansion would have a
finite radius of convergence.
\\ \\
The zero-dimensional theory can be used for counting planar
diagrams. It can be solved explicitly, so that the generating function
for the number of diagrams with given 3-vertices and 4-vertices, can
be derived exactly. This can be done for various kinds of Feynman
diagrams. We end with some remarks about planar renormalization.}

%\Endnarrower \hugeskip
%\newsect{1. Introduction. The Large $N$ Sigma Model.}
\section{Introduction: The Large $N$ Sigma Model}

A simple example of a theory where new and interesting information
can be obtained by studying it in the limit where the number $N$
of fundamental field degrees of freedom becomes large, is the
Large $N$ Sigma Model\ref1. Let $\f_i(x)$ be a set of $N$ scalar
fields, $i=1,\cdots,\,N$, and consider the Lagrangian of a
renormalizable theory:
$$\LL=-\half\sum_{i=1}^N\left[(\pa_\m\f_i)^2+m^2\f_i^2\right]-
\fract18\l\left(\sum_{i=1}^N\f_i^2\right)^2\ ,\eqno(1)$$ where
$\l$ is a coupling constant, usually taken to be positive.

\def\lift#1#2{\setbox0=\hbox{$#2$}\raise#1\box0} % bijv: lift{1em}{\sigma}

The Feynman diagrams of the theory consist of propagators,
$$\epsffile{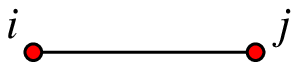}\qquad {\d_{ij}\over k^2+m^2-i\e}\
,\eqno(2)$$ to be connected by vertices:
$$\epsffile{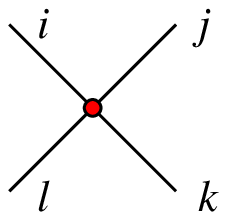}\qquad\lift{1.8em}{-\l(\d_{ij}\d_{k\ell}+
\d_{ik}\d_{j\ell}+\d_{i\ell}\d_{jk})\ .} \eqno(3)$$ We write this
vertex as the sum of three vertices, to be indicated the following
way: $$ \lift{-2.6em} {\epsffile{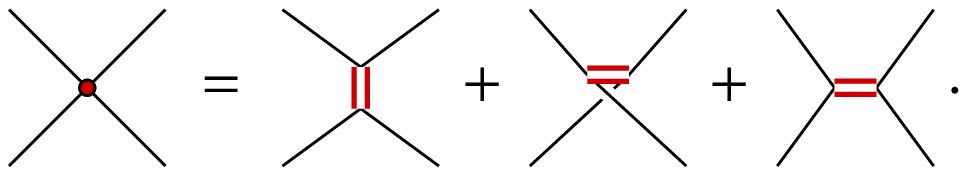}} \eqno(4)$$

Rewriting the Lagrangian~(1), using an auxiliary field $A(x)$,
as $$\LL=-\half\sum_{i=1}^N\left[(\pa_\m\f_i)^2\right]
+\fract1{2\l} A^2-\half(A+m^2)\left(\sum_{i=1}^N\f_i^2\right)\
,\eqno(5)$$ we get the same Feynman rules [since integrating out
$A$ yields back Eq.~(1)], but now the double line is the $A$
propagator, $-\l$, and the $A$ lines form 3-vertices with the $\f$
lines. It is now relatively easy to integrate out the field $\f$
in (5). We must remember that $A$ is $x$ dependent, so we do
this perturbatively in the $A$ field. This way, we get effective
interactions in the $A$ field, the interaction being described by
`vertices' formed by closed $\f$ lines, see Fig.~1. Notice that
the set of Feynman rules obtained this way is easily seen to be
the same as the ones generated by Eqs.~(1)-(4), as of course
it should be.
%
%\begin{figure}
$$\lift{-3em}{\epsffile{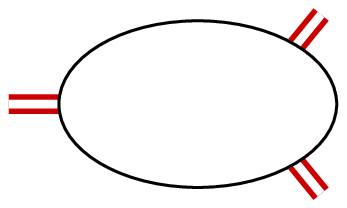}}\footnotesize
\matrix{\hbox{Figure 1. Effective vertex for $A$ field.}}$$
%\end{figure}
%
Since (5) is only quadratic in $\f$ the effective vertices only
consist of single $\f$ loops, and so they are easy to calculate.

{\it Since there are $N$ fields $\f$, each of these effective
vertices go with a factor $N$.} It is now instructive to replace
the field $A$ by $\tilde{A}/\sqrt N$, which amounts to a
propagator $-\l N$ for $\tilde A$, whereas the $\n$-point
effective vertices each receive an extra factor $N^{-\n/2}$.
Together with the factor $N$ mentioned above, we find that the
effective 2 point vertex goes with a factor $N^0$ (so it is
$N$-independent), the 3-vertices go with $N^{-1/2}$, etc. The {\it
tadpole diagram\/} (the effective vertex with $\n=1$) must be
assumed to vanish, after application of a normal ordering
procedure. It would be proportional to $N^{+1/2}$.

If now we choose the limit $N\ra\infty$, while $\tilde\l=\l N$ is
kept fixed, then we see that a perturbative theory emerges, where
all effective vertices with $\n$ external lines are suppressed by
factors $N^{1-\n/2}$, so that $1/\sqrt N$ acts as a coupling
constant. Only the 2-point vertex is non-perturbative, so it has
to be computed exactly, and added to the `bare' term $-\tilde\l$
to obtain the $\tilde A$ propagator. Fortunately, the 2-point
vertex correction $F(m,q)$, where $q$ is the momentum of the
$\tilde A$ field and $m$ is the $\f$ mass, is easy to compute. In
$n$ spacetime dimensions,
\begin{eqnarray} \setcounter{equation}{6}
F(m,q) & = & \half\, (2\pi)^{-4}\int\dd^nk{1\over(k^2+m^2-i\e)
[(k+q)^2+m^2-i\e]} \nonumber \\ & = &C(n)\int_0^1\dd
x[m^2+q^2x(1-x)]^{\fract n2-2}\ ,
\end{eqnarray}
where $C(n)=\half(4\pi)^{-n/2}\,\G(2-\fract n2)$.

This has to be added to the bare $\tilde A$ propagator term, so
the series of diagrams contributing to the dressed propagator
amounts to: $$\lift{-1.16em}{\epsffile{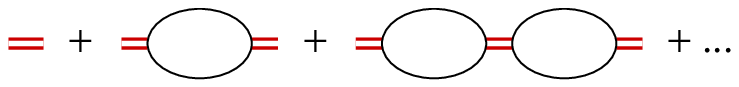}}\quad =
{-\tilde\l\over 1+\tilde\l F(m,q)}\ . \eqno(7)$$ This is the exact
propagator to lowest order in $1/N$. The effective Feynman
diagrams describing the $1/N$ corrections are sketched in Fig.~2.
In many respects, the $1/N$ expansion is like the coupling
constant expansion in a field theory. Thus, the effective
propagators in diagrams such as Fig.~2 can form closed loops.
\begin{figure} \setcounter{figure}{1}
\begin{center}
{\epsffile{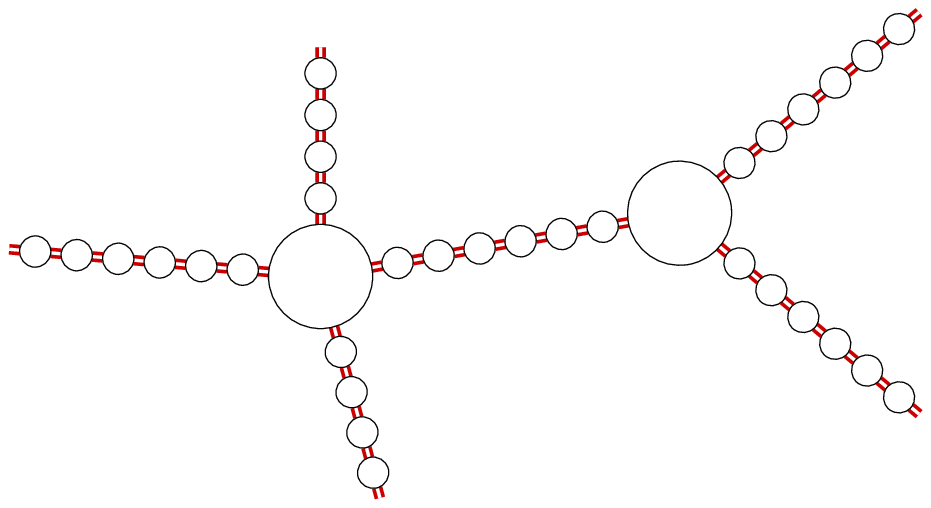}}
\caption{Effective Feynman graphs for the $N\ra\infty$
Sigma Model.}
\end{center}
\end{figure}

At $n=4$, the function $C(n)$ has a pole, which has to be absorbed
into a redefinition of $1/\tilde\l$, as usual. At $n=6$ or higher,
there are divergences which require more troublesome subtractions,
since there the theory is non-renormalizable. It is instructive to
look at the $q$ dependence of this effective propagator. At $n<4$,
this propagator is regular at high $\pm q^2$, since $F$ tends to zero.
$F$ is complex if $q^2<-4m^2$. If $m$ is low enough, and/or $\tilde\l$
large enough, then a new pole may emerge: an (unstable) bound state of
two $\f$ particles. If, however, $n\ge4$, the function $F$ diverges at
{\it large} $\pm q^2$. Note that the {\it sign\/} of $F$ after the
required subtraction flips, so that for large $q^2$ $F$ behaves as
$$F(m,q^2)\ra -\fract1{32\pi^2}\log q^2\ .\eqno(8)$$ Consequently, in
Eq.~(7) a pole at some large {\it positive\/} value of $q^2$ is
inevitable. Thus, we have a tachyonic pole in the propagator: the
theory has a Landau ghost. This ghost is directly related to the fact
that the theory is not asymptotically free. At very large momenta, the
effective quartic coupling is negative, which implies a fundamental
instability of the theory.  This is an important lesson from the study
of the $1/N$ expansion of the Sigma Model: lack of asymptotic freedom
in the ultraviolet makes the theory inconsistent. There is no tendency
to go to an ultraviolet fixed point. The theory does not show any
inclination to ``cure itself." In the literature, we often see the
belief that models with a decent-looking Lagrangian are well-defined,
``if only we knew how to do the mathematics." Here, we happen to be
able to do the mathematics, and find the model to be inconsistent at
$n\ge4$ due to its bad ultraviolet behaviour, just like any other
non-renormalizable system.

%\newsect{2. $1/N$ for Gauge Theories.}
\section{$1/N$ for Gauge Theories}
Many gauge theories are asymptotically free\ref2, so here a large
$N$ expansion could be much more promising. The large $N$
expansion for gauge theories can be formulated for all large Lie
groups, and the final results are all very similar. Let us first
consider the gauge group $U(N)$. The vector fields are Hermitean
matrices $A_{\m j}^{\ i}$, where both indices $i$ and $j$ run from
1 to $N$. Writing
\begin{equation} \setcounter{equation}{9}
F_{\m\n j}^{\ \ i}=\pa_\m A_{\n j}^{\
i}-\pa_\n A_{\m j}^{\ i}+ig\,[A_\m,\,A_\n]^i_j\,,
\end{equation}
we have
\begin{eqnarray}
\LL & = & -\quart{\rm Tr}
(F_{\m\n}F_{\m\n})-{\overline\j}(i\g D+m_\j)\j\ \nonumber \\ & = & \
-{\rm Tr}\Big(\half(\pa_\m A_\n)^2-\half\pa_\m A_\n\pa_\n
A_\m+ig\,\pa_\m A_\n[A_\m,\,A_\n]-\quart g^2\,
[A_\m,\,A_\n]^2\Big) \nonumber \\ && -
{\overline\j}(i\g_\m\pa_\m+ig\g_\m A_\m+m_\j)\j
\,.
\end{eqnarray}
A transparent notation for the Feynman diagrams\ref3 is one where the
propagators are represented by double lines, each line indicating how
the index is transported, see Fig.~3. The mathematical details of
these rules depend on the gauge fixing procedure, which we shall not
further expand upon.
\begin{figure}
\begin{center}
\epsffile{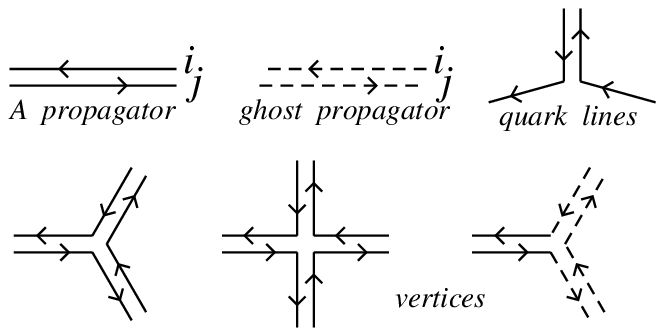}
\end{center}
\caption{Feynman rules for $U(N)$ gauge theories.}
\end{figure}

Everywhere in a Feynman diagram where an index loop closes, we get
a factor $N$ from counting the values that the index can have.
This is the only way the dimension $N$ of the gauge group enters.
Now consider a general diagram with $V_3$ 3-point vertices, $V_4$
4-point vertices, $P$ propagators and $F$ index loops. Assume,
temporarily, no external lines (a gauge-invariant coupling to an
external source can for instance be introduced by means of a 2
quark vertex insertion, $J{\overline\j} \j$; this then is a 2
point vertex, counted by $V_2$).  Write $V=V_2+V_3+V_4$.

First, by drawing dots at both ends of every propagator, we see
immediately by counting dots that $$2P=\sum_n nV_n\,.\eqno(11)$$ We
now turn the diagram into a polyhedron by defining every index loop to
be a facet. The facets form a polyhedron if there are no quark
loops. Then, we have Euler's theorem for polyhedra:
$$F-P+V=2-2H\,.\eqno(12)$$ Here, $H$ is the {\it genus\/}, or number
of `holes' in the polyhedron. This theorem is easily proven by
induction. Since all 3-point vertices in Fig.~3 come with a factor
$g$, and all 4-point vertices with a factor $g^2$, the $g$ and $N$
dependence of the amplitude is seen to be\ref3
$$g^{V_3+2V_4}N^F=g^{2P-2V}N^F=(g^2N)^{F+2H-2}N^{2-2H}\,.\eqno(13)$$
In the limit where $N\ra\infty$, while $\tilde g^2=g^2N$ is kept
fixed, we see that the $N$ dependence of the diagrams is $N^{2-2H}$,
so that diagrams without holes ($H=0$) dominate.

If quarks are present, we simply declare the quark loops to be
facets as well, but they lack a factor $N$. Therefore, we also get
a factor $N^{-Q}$, where $Q$ is the number of quark loops. Thus,
the $1/N$ expansion is found to be an expansion with respect to
the number of quark loops and the number of holes in the
polyhedra. In many respects, this expansion resembles an expansion
of a string theory in powers of the string coupling constant.

For other gauge groups, the result is very similar. If we take
$SU(N)$ instead of $U(N)$, we are essentially subtracting a $U(1)$
component, whose coupling is $g=\tilde g/\sqrt N$, so even at
large $\tilde g$, the $U(1)$ subtractions are still perturbative.
For $SO(N)$, we have an additional reality constraint on the
fields $A$: $$A_{\m j}^{\ i}=-A_{\m i}^{\ j}={A_{\m j}^{\ i}}^*\
  .\eqno(14)$$ This means that the arrows disappear from the
double line propagators in Fig.~2. The surfaces of the planar
diagrams are now non-oriented, so that non-orientable diagrams are
allowed: Klein bottles and M\"obius strips. They, however, have
$H>0$ or $Q>0$, and hence they are also suppressed at large $N$,
so that the dominating diagrams are essentially the same as the
ones for $U(N)$ and $SU(N)$.

%\newsect{3. $1/N$ for Theories on a Lattice.}
\section{$1/N$ for Theories on a Lattice.}
\def\ul{\underline}
It is instructive to consider the $1/N$ expansion for lattice
gauge theories\ref4. Consider a lattice with lattice sites,
connected by lattice links. The gauge fields live on the links,
the quark fields live on the sites. On a link with length $a$,
connecting the sites $x$ and $x+e_\m$ (where $e_\m$ is a unit
vector of length $a$ on the lattice), we write
$$U_\m(x)\deff\exp(ia\,g\, A_\m)\,.\eqno(15)$$ The
gauge-invariant Wilson action (in $n$ dimensions) is
$$S={a^{n-4}\over 2 g^2}\sum_{x,\m,\n}{\rm
Tr}\left[U_\m(x)U_\n(x+e_\m)U_\m^\dagger(x+e_\n)U_\n^\dagger(x)
\right]\,.\eqno(16)$$ The two indices of the field variable
$U_{\m\,i}^{\ j}(x)$ refer to gauge transformations in $x$ and
$x+e_\m$, respectively, so these are two different indices, to be
distinguished by underlining one of them. Since $U(x)$ is unitary,
we have $$U_i^{\,\ul j}U_{\ul j}^{\dagger k}=\d_i^k\ ;\qquad
U_{\ul i}^{\dagger j}U_j^{\,\ul k}=\d_{\ul i}^{\,\ul k}\
,\eqno(17)$$ from which we derive that the `functional' integral
over the $U$ values will yield\ref5: $$\bra U_i^{\,\ul j}U_{\ul
k}^{\dagger\ell}\ket={1\over N}\d_i^\ell\d_{\ul k}^{\,\ul j} \ ,
\eqno(18)$$ [simply by observing that (18) is the only term with
the required gauge invariance, so all that has to be determined is
the prefactor]. Similarly, we have only four possible terms for
the quartic averaging expression. Two of these turn out to be
leading: $$\bra U_i^{\,\ul j}U_{\ul k}^{\,\ell}U^{\,\ul
n}_{\lift{-.1em}{\scriptstyle m}} U_{\ul p}^{\,q} \ket={1\over
N^2}\left(\d_i^\ell\d_{\ul k}^{\,\ul
j}\d_m^{\lift{.1em}{\scriptstyle q}}\d_{\ul p}^{\,\ul
n}+{\textstyle{ \ell\leftrightarrow q\atop\ul k\leftrightarrow\ul
p}} \right)+\OO(N^{-3})\ .\eqno(19)$$ The effect of these
expressions is that when we do the functional integral, simply by
inserting the complete expansion of the exponential of the action
$S$, at every link every factor $U$ has to be paired with a factor
$U^\dagger$. Every such pair produces a factor $1/N$.  As in the
continuum case, we construct a polyhedron from the elementary
plaquettes. Let $F$ be the number of plaquettes (the number of
terms of $S$ in our `diagram', coming from expanding $e^{iS}$),
let $P$ be the number of pairings, and $V$ be the number of
indices that end up being summed. We find a factor $1/N$ for each
pairing, a factor $N$ for each index, which amounts to a factor
$N$ for each point on our polyhedron, and a factor $1/g^2$ for
every plaquette (apart from a harmless numerical coefficient).

All in all, the $N$ and $g$ dependence arises as a factor
$$N^V(1/N)^P(g^2)^{-F}=(g^2N)^{-F}N^{V-P+F}=({\tilde
g}^2)^{-F}N^{2-2H}\ ,\eqno(20)$$ where again Euler was applied, and
${\tilde g}^2=g^2N$. So again we find that the diagrams only depend on
$\tilde g$, and the $N$ dependence of the diagrams is as in the
continuum theory. This result is not quite trivial, since we are
looking at a different corner of the theory, and the transition from
the small $\tilde g$ to the large $\tilde g$ limit need not commute
with the large $N$ limit, but apparently it does.

%\newsect{4. Divergence of Perturbation Expansion.}
\section{Divergence of Perturbation Expansion}
Gauge theories can now be expanded in terms of several parameters:
we have the $\tl g$ expansion, the $1/\tl g$ expansion and the
$1/N$ expansion. Of these, only $1/N$ acts as a genuine interaction
parameter, describing the strength of the interactions among
mesons. Physical arguments would consequently suggest that the
$1/N$ expansion produces terms of the form $a_k\,N^{-k}$ where
$a_k$ will diverge faster than exponentially for large $k$ values,
typically containing $k!$ terms. In contrast, the $\tl g$
expansion generates only planar diagrams. In principle, one could
expect better convergence for this expansion. Indeed, the {\it
number\/} $R_k$ of planar diagrams with $k$ loops will be bounded
by an expression of the form $C^k$ for some finite constant $C$.

Unfortunately, even the $\tl g$ expansion will generate factors
$k!$, not due to the number of diagrams being large, but because
of renormalization effects. These effects already occur in the
large $N$ sigma model of Sect.~1. Consider the $\tl\l$ expansion
of the propagators of Figure~2. The expansion term going as
$\tl\l^{k+1}$ contains exactly $k$ loops. Each of these loops
generates an expression $F(m,q)$, which, for large $q$, diverges
roughly like $\log(q^2+m^2)$. Imagine this propagator itself being
part of a closed loop, and imagine that a sufficient number of
subtractions has been carried out in order to make the integral
convergent, so that some renormalized amplitude results. We see
that the $k^{\rm th}$ term requires the calculation of an integral
of the form $$\int\dd^4q{\big[\log(q^2+m^2)\big]^k\over
(q^2+m^2)^3}\ ;\eqno(21)$$ Writing $q^2+m^2\approx m^2e^x$, we see
that integrals of the form $$\int_0^\infty\dd x\,x^k e^{-x}=
k!\eqno(22)$$ arise.

Such $k!$ terms being very similar to the ones generated by instanton
effects in quantum field theory, this effect was dubbed
`renormalon'. The renormalon is closely associated with the Landau
ghost. Indeed, replacing the coupling constant $\tl\l$ by a running
coupling constant, we notice that the Landau ghost mass increases
exponentially with the inverse coupling constant: $$m_{\rm gh} \propto
e^{16\pi^2/\tl\l}\,.\eqno(23)$$

In some very special cases, the Landau ghost can be avoided
altogether. This is if the theory is both asymptotically free in
the ultraviolet and infrared convergent due to mass terms\ref6.
Such theories can be constructed, making judicious use of gauge
vector, scalar {\it and\/} spinor fields. Integrals of the type
(22) do still occur in that case, but the perturbation expansion
can then be Borel resummed\ref6.

QCD is far more divergent than that. Here, we have no mass term to
curb the infrared divergences. Instanton effects are to be
expected at finite $N$, and possibly even at infinite $N$, and
infrared renormalons invalidate simple attempts to prove
convergence of Borel resummation\ref7. Simple physical
considerations, however, could be used to argue that unique
procedures should exist to resum the planar diagrams.

%\newsect{5. Counting Planar Diagrams.}
\section{Counting Planar Diagrams}

Doing such resummations analytically is way beyond our present
capabilities. What we can do is carry out such summations in a
simple, zero-dimensional theory. This result will give us an
analytic method to {\it count\/} planar diagrams.  The `field
theory' is described by the action\ref{8,9} $$S(M)={\rm
Tr}\left(-\ffract12 M^2+\ffract13 g M^3+\ffract14 \l M^4\right),
\eqno(24)$$ where $M$ is an $N\times N$-dimensional matrix, in
the limit $$N\ra\infty\ ,\qquad g,\,\l\ra 0 ,\qquad N\,g^2=\tl
g^2\ \hbox{\ and\ }\ N\l=\tl\l\ \hbox{\ fixed.}\eqno(25)$$

Henceforth, the tilde ($\tl{\,}$) will be omitted. Precise
calculations of the generating functions for the number of planar
diagrams, with various types of restrictions on them, were carried
out in Ref.~\ref{10}.  Different cases were studied, all related
to one another by exact equations. Part of that paper is based on
pioneering work by Koplik, Neveu and Nussinov\ref8--- which in
turn makes use of earlier work by Tutte\ref9--- and on the work by
Br\'ezin, Itzykson, Parisi and Zuber\ref{11}.  The latter apply
matrix theory to do the integration with the action~(1). It is
instructive instead to work directly from the functional
equations, as these will be easier to handle in the QCD case, and
they are also more transparent in diagrammatic approaches.  The
relations are read off directly from the diagrams.

A delicate problem then is the choice of boundary conditions for
these equations.  They can be derived by carefully considering the
holomorphic structure that the generating functions are required
to have.  Once this is understood, a fundamental solution is
obtained for the generating function describing {\it the numbers
of all planar diagrams for all multiparticle connected Green
functions, with given numbers $V_3$ of three-point vertices, $V_4$
four-point vertices, and $E$ external lines}.

We denote the number of all connected planar diagrams by
$N_{(E,V_3,V_4)}$. The generating function $F(z,g,\l)$ is defined
by $$F(z,g,\l)\equiv\sum_{ E=1,\atop
V_3=0,\,V_4=0}^{\infty,\,\infty,\,\infty}
z^{E-1}\,g^{V_3}\,\l^{V_4}\,N_{(E-1,V_3,V_4)}\ .\eqno(26)$$ Here,
it is for technical reasons that we start counting at $E=1$.

The recursion equation for the generating function $F\deff
F(z,g,\l)$, is
\begin{eqnarray} \setcounter{equation}{27}
& F & =\  z +\ g\Big(F^2\ +\ {F-F_{(0)}\over z}\Big)\ \nonumber \\
& + & \l\left(F^3\ +\  2F{ F-F_{(0)} \over z}\ +\
{F-F_{(0)}-F_{(1)}z\over z^2} \ +\  F_{(0)}{F-F_{(0)} \over
z}\right) .{\ }
\end{eqnarray}
Here, $F_{(0)}=F(0,g,\l)$ is the tadpole diagram ($E=1$), and
$F_{(1)}=$ ${\pa\over\pa z}F(z,g,\l)|_{z=0}$ is the propagator
($E=2$). Written diagrammatically, we have\br
%\indent
\epsffile{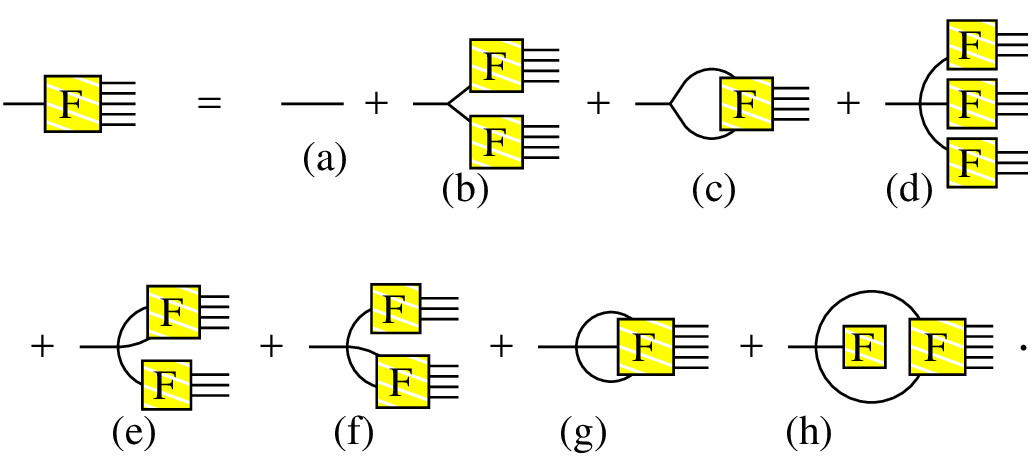}~\hfill~\lift{4em}{(28)} \vspace{1ex}
%\hfill \lift{3em}{(28)}

\ni Note that inside the 1-loop terms the tadpole diagrams had to
be removed by hand, and in the 2-loop term the propagator had to
be removed.  Such equations can be solved! There are three
methods:
\begin{list}
{\it\roman{L1}.}{\usecounter{L1}}
\item The first terms in these lists of expressions can be
found numerically. We could then try to guess the generic
expressions for $N_{(E-1,V_3,V_4)}$, and check these expressions
in the closed equation~(27) or (28). This works only in the very
simplest of cases.
\item One can study the holomorphic
structure of $F_0(z,g,\l)$ near the origin, $g=\l=0$. Demanding the
correct non-singular behaviour at the origin provides the unique
boundary condition.
\item One can return to the definition of the generating
functions using the integral expression for large matrices, and take
the limit for the infinite matrices $M$. The integrand contains only
one matrix, which can be conveniently diagonalized (in case of two or
more fields, or in dimensions higher than zero, simultaneous
diagonalization of the matrices $M_a(x)$ is impossible, which leads to
considerable complications). This turns out to be the most elegant
method.
\end{list}
After some work, we found that all approaches
lead to consistent results. One first derives the values of $F_{(0)}$
and $F_{(1)}$, after which the values for $F(z,g,\l)$ follow by
formally solving Eq.~(27).

The expressions found for $F(0,g,\l)$ turn out to be singular on a
curve in the $g$-$\l$ plane. This gives us the critical values of
$g$ and $\l$, in terms of a parameter $t$ that parametrizes this
curve:
\def\tta{6-6\,t-3\,t^2+2\,t^3}
\def\ttb{12-4\,t^3+t^4}
\def\ttc{2+2\,t-t^2}
\begin{eqnarray} \setcounter{equation}{29}
g_0^c & = & 2\,t^2\,(3-t)(\tta)^{1/2}\,(\ttb)^{-3/2}\,;\vphantom{\Big|_p}
\nonumber \\
\l_0^c & = & (\ttc)\,(\tta)\,(\ttb)^{-2}\,.
\end{eqnarray}

All real values for $g$ are allowed, but $\l$ must be positive.
The curve defined by Eqs.~(29) in the $g$-$\l$ plane then looks as
in Fig.~3. Note that it is the $g$ dependence, not the $g^2$
dependence that we are looking at. The cusp at $g=0$ is a genuine
singularity of the curve (29).
\begin{figure}
\begin{center}
\epsffile{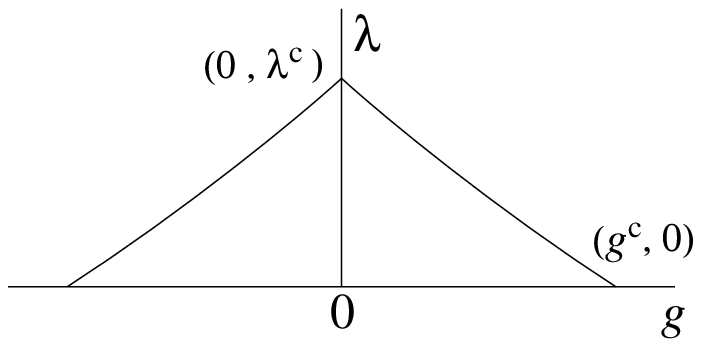}
\end{center}
\caption{Line of critical $\l$ and $g$ values to a high precision.  The
lines are not exactly straight, and in fact form a single curve with a
cusp singularity. The corners are given by the values
$\l^c=\fract1{12}$ and $g^c=1/\sqrt{12\sqrt3}$.}
\end{figure}

Next, we can make restrictions on the occurrence of insertions
within the diagrams. This we indicate by using a subscript $a$,
taking values between 0 and 5:  \prefbreak{1cm}
\begin{eqnarray}
a=0: && \quad\hbox{no further restrictions.} \\
a=1: && \quad\hbox{no tadpoles:}\epsffile{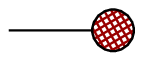} \\
a=2: && \quad\hbox{no tadpoles and no seagulls:}
\epsffile{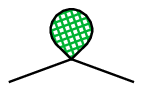}\\
a=3: && \quad\hbox{also no self-energies:}\qquad \,\epsffile{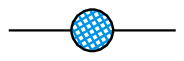}\\
a=4: && \quad\hbox{also no dressed 3-vertices:}\epsffile{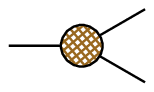}\\
a=5: && \quad\hbox{also no dressed 4-vertices:}\epsffile{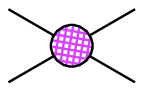}
\end{eqnarray}
\def\ttd{144 - 288\,t + 192\,{t^3} - 168\,{t^4} - 48\,{t^5} +
100\,{t^6} - 8\,{t^7} - 17\,{t^8} + 4\,{t^9}}
\def\tte{ 336 - 672\,t + 384\,{t^2} + 256\,{t^3} - 712\,{t^4}
 + 48\,{t^5} + 308\,{t^6} - 56\,{t^7} - 45\,{t^8} + 12\,{t^9}}
\def\ttf{ 24 - 72\,t - 12\,{t^2} + 56\,{t^3} - 10\,{t^4}
 - 10\,{t^5} + 3\,{t^6}}

In each of these cases, there is a domain in the $g$-$\l$ plane
where the zero-dimensional theory converges. The location of the
boundaries of these domains then tells us how the number of
diagrams of the specific types increase at high orders. For
example, the last curve, of the case $a=5$, is given by the
parametric equations
\begin{eqnarray}
g_5^c & = & {2\,t^2\,(6-t^2)^2(6-3\,t^2+t^3)\over(\ttf)^{3/2}}
\, , \vphantom{\bigg|_p} \nonumber \\
 \l_5^c & = & \big(3456 - 31104\,t + 25920\,{t^2} + 55296\,{t^3} -
 180576\,{t^4} - 44064\,{t^5} \, \nonumber \\ & &\quad + \, 247824\,{t^6}
 - 27936\,{t^7} - 147672\,{t^8} + 52760\,{t^9}+ 38076\,{t^{10}} -
  24432\,{t^{11}}\, \nonumber \\
& & \quad - \,1762\,{t^{12}}+4506\,{t^{13}}
 - 921\,{t^{14}} - 202\,{t^{15}} + 114\,{t^{16}} - 18\,{t^{17}}
 + {t^{18}}\big)\, \nonumber \\ &
   &\vphantom{\big)}\qquad\qquad \times (\ttf)^{-3}\, .
\end{eqnarray}
\begin{figure}
\begin{center}
\epsffile{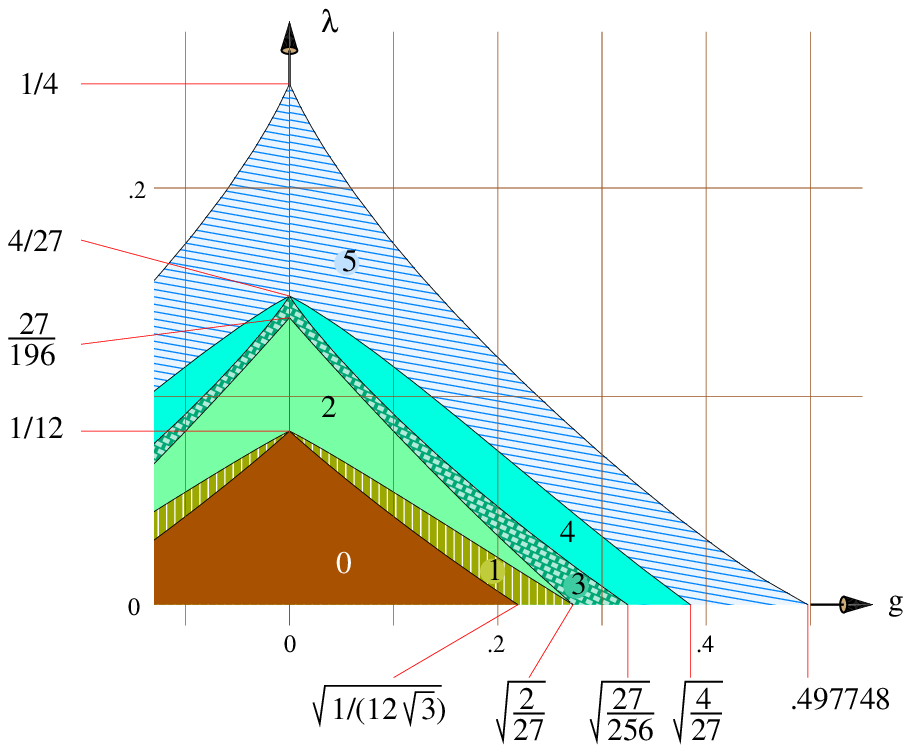}
\end{center}
\caption{Critical regions after imposing different constraints.}
\end{figure}

These, as well as the other expressions that were derived in
Ref.~\ref{10}, are all exact. The domains of convergence for the
various cases are displayed in Fig.~5. The last boundary, that of
region \# 5, is the one given by Eqs.~(36). Figures~4 and~5 show
the actual domains with great precision. All curves have cusps at
$g=0$. Many of our results were obtained and/or checked using the
computer program {\sl Mathematica}.

%\newsect{6. The $g^2N$ expansion.}
\section{The $g^2 N$ Expansion}
We conclude with a short discussion on constructing perturbative
planar QCD amplitudes at all orders without divergences.

A renormalization scheme that appears to be not so well-known\ref6
can be set up in the following way.  It works particularly well
for planar field theories in 4 space-time dimensions, but it also
applies to several other quantum field theories, notably QED.
Presumably, QCD with $N_c=3$ can also be covered along these
lines, but some technical details have not been worked out to my
knowledge. Our scheme consists of first collecting all
one-particle irreducible 2-, 3- and 4-point diagrams, and formally
considering the non-local quartic effective action generated by
these diagrams.

Next, consider all diagrams using the Feynman rules derived from
this action, but {\it with the limitation that only those diagrams
that are absolutely ultraviolet convergent are included}.  No
superficially ultraviolet divergent subgraph is accepted.  To be
precise, we omit all diagrams with 4 or less external lines, as
well as all diagrams containing any non-trivial irreducible
subdiagram with 4 or less external lines. They are the diagrams of
type 5 of the previous section.

Clearly, one expects that the 1PI subgraphs with 4 or less
external lines have already been taken care of by our use of the
quartic effective action instead of the bare Lagrangian.  The
important issue to be addresed is, whether the counting of all
diagrams was done correctly so as to obtain the required physical
amplitudes.  But this is not difficult to prove:

$\underline{\hbox{Theorem:}}$ The above procedure correctly
reproduces the complete amplitudes for the original theory.  No
diagrams are over-counted or under-counted.

The proof of the theorem is to be found in Refs\ref{6,10}. One
deduces that\br {\it if all one-particle irreducible 2-, 3-, and
4-point vertices are known, all amplitudes can be derived without
encountering any divergent (sub-)graph(s).}

Subsequently, what has to be done to complete a perturbative
computational scheme, is to establish an algorithm to compute the
irreducible 2-, 3-, and 4-point vertices (and, if they occur, the
tadpole diagrams as well). Actually, this is simple. Consider a
4-point 1PI diagram $\G(p_1,p_2,p_3,p_4)$. Here $p_\m$ are the
external momenta, and $p_1+p_2+p_3+p_4=0$. Now consider the
difference $$\G(p_1+k,p_2-k,p_3,p_4)-\G(p_1,p_2,p_3,p_4)\equiv
k_\m\D^\m(p_1, \underline{k},p_2-k,p_3,p_4) \,.\eqno(37)$$ The
underlining here refers to the fact that, in the function $\D^\m$
the external line with momentum $k$ follows distinct Feynman
rules.

If we follow a path inside the diagram, we can consider the entire
expression for $\D^\m$ as being built from expressions containing
differences. For instance, in the propagators:
$${1\over(p+k)^2+m^2}-{1\over p^2+m^2}\ =\ {k_\m\,(-2p-k)^\m\over
\big[(p+k)^2+m^2\big](p^2+m^2)}\,,\eqno(38)$$ or else, in the
3-vertices: $$(p+k)_\n\ -\ p_\n\ =\ k_\m\, \d^\m_{\,\n}\
 .\eqno(39)$$ We notice that the expressions for $\D^\m$ are all
(superficially) ultraviolet convergent! Actually, one may set up
unambiguous Feynman rules for $\D^\m(p_1,
\underline{k},p_2,-kp_3,p_4)$ and observe that this amplitude
exactly behaves as a 5-point diagram, hence it is (superficially)
convergent.

For the 3-point and the 2-point diagrams, one can do exactly the
same thing by differentiating more than once. In practice, what
one finds is that there is a set of rules containing fundamental
irreducible 2-, 3- and 4-point vertices, and in addition rules to
determine their differences at different values of their momenta.
The complete procedure thus leads to the following situation.

\def\uunderl#1{\underline{\underline{#1}}}

We start by postulating the so-called `primary vertex functions'.
These are not only the irreducible 2-point functions
$\G_{[2]}(p,-p)$, the irreducible 3-point functions
$\G_{[3]}(p_1,p_2,-p_1-p_2)$ and the irreducible 4-point functions
$\G_{[4]}(p_1,\dots,p_4)$, but also, in addition, the difference
functions $\D^\m_{[2]}(p,\underline{k},-p-k)$ and
$\D^\m_{[3]}(p_1,\underline{k} ,p_2-k,-p_1-p_2)$, and finally the
functions $U_{[2]}$, obtained by differentiating $\D^\m_{[2]}$
once more:
$$\matrix{\D_{[2]}^\m(p_1+q,p_2-q,-p_1-p_2)-\D_{[2]}^\m(p_1,p_2,-p_1-p_2)\equiv\cr
\hfil\hfil\vphantom{\big]^k} q_\n
U_{[2]}^{\m\n}(p_1-q,\uunderl{q},p_2,-p_1-p_2)\,,}\eqno(40)$$
where one of the other external lines, $p_1$ or $p_2$, is
underlined. The double underlining is here to denote that the two
entries are to be treated distinctly (because of the factor
$k_\m$, the functions $U_{[2]}^{\m\n}$ are not symmetric under
interchange of $k$ and $q$).

These primary vertex functions are derived by first considering
the differences for $\G_{[4]}$, $\D^\m_{[3]}$ and $U^{\m\n}_{[2]}$
at two different sets of external momenta. These expressions are
handled as if they were irreducible 5-point diagrams. These are
expanded in terms of planar diagrams, where all irreducible
subraphs of 4 or less external lines are bundled to form the
primary vertex functions. At one of the the edges of such a
diagram, we then encounter one of the functions $\D^\m$ or
$U^{\m\n}$.

This way, we arrive at difference equations for the primary vertices,
with on the r.h.s.\ again the primary vertices. The primary vertices
$\G_{[3]}$, $\D^\m_{[2]}$ and $\G_{[2]}$ are then obtained by
integrating equations such as~(37) with respect to the external
momentum $k$. The {\it integration constants} will be associated with
the values of the vertices and propagators in the far ultraviolet
region, where the theory is asymptotically classical. This completes
the procedure to obtain all amplitudes by iteration.

Technical implementation of our scheme requires that in all
diagrams, an unambiguous path can be defined from one external
line to another. In QED, one may use the paths defined by the
electron lines. In a planar theory, one may define the paths to
run along the edges of a diagram.

Three remarks are in order:
\begin{list}
{\it\roman{L1}.}{\usecounter{L1}}
\item The procedure is effectively a renormalization group procedure.
The functions $\D^\m$ and $U^{\m\n}$ play the role of beta
functions.
\item The procedure is essentially still perturbative, since the planar
diagrams must still be summed. Our beta functions are free of
ultraviolet divergences, but the summation over planar diagrams may
well diverge.
\item The procedure only works if the integrations do not lead to
clashes. This implies that it is not to be viewed as a substitute for
regularization procedures such as dimensional regularization.  We
still need dimensional regularization if we want to {\it prove\/} that
the method is unambiguous, which, of course, it is in the case of
planar QCD.
\end{list}

%\bye
\end{document}